\documentclass[nocccite]{cc}
\usepackage[english]{babel}
\usepackage{graphicx}
\newcommand {\hide}[1]{}
\newcommand {\s}        {\mbox{\rm sign}}
\newcommand {\D}     {\mbox{\rm D}}
\newcommand {\R}     {\mbox{\rm R}}

\newcommand {\C}     {\mbox{\rm C}}

\newcommand {\Z}  {{\mathbb Z}}
 
\newcommand {\Q}         {{\mathbb Q}}

\newcommand {\ZZ} {{\rm Z}}
\newcommand {\RR} {{\cal R}}

\newcommand {\eps} {{\varepsilon}}
\newcommand {\E} {{\rm Ext}}

\def\addots{\mathinner{\mkern1mu
\raise1pt\vbox{\kern7pt\hbox{.}}
\mkern2mu\raise4pt\hbox{.}\mkern2mu
\raise7pt\hbox{.}\mkern1mu}}

\newcommand {\SIGN} {{\rm Sign}}

\newcommand{\coucou}[1]{\ifvmode\else\marginpar[\hfill$\rhd$]{$\lhd$}\fi
                       $\langle$\textsc{#1}$\rangle$}

\newcommand{\basu}[1]{}
\newcommand {\rp}[1]{}

\contact{saugata@math.gatech.edu}
\received{March 21, 2005.}
\title{
Efficient algorithm for computing the Euler-Poincar\'e characteristic
of a semi-algebraic set defined by few quadratic inequalities
}
\titlehead{Computing Euler-Poincar\'e Characteristics}
\author{Saugata Basu \\
School of Mathematics,\\
Georgia Institute of Technology,\\
Atlanta, GA 30332, U.S.A.\\
\email{saugata@math.gatech.edu}
}
\input{diagrams}
\newarrow{Line}{.}{.}{.}{.}{.}

\begin{abstract}
We present an algorithm which takes as input a closed semi-algebraic set,
$S \subset \R^k$, defined by 
\[
P_1 \leq 0, \ldots, P_\ell \leq 0,
P_i \in \R[X_1,\ldots,X_k], \deg(P_i) \leq 2,
\]
and computes the Euler-Poincar\'e characteristic of $S$. The complexity of the
algorithm is $k^{O(\ell)}$.
\end{abstract}


\begin{subject}
2000 Mathematics Subject Classification 14P10, 14P25
\end{subject}

\begin{document}
\section{Introduction}
\label{sec:intro}
Let $\R$ be a real closed field and let $S \subset \R^k$ be a basic
semi-algebraic set defined by 
$P_1 \leq  0, \ldots, P_\ell \leq 0,$ 
with $P_i \in \R[X_1,\ldots,X_k], \deg(P_i) \leq 2, \; 1 \leq i \leq \ell$. 
It is known \cite{Barvinok,B03}
that the sum of the Betti numbers of $S$
(and hence also its Euler-Poincar\'e characteristic) is bounded by
$k^{O(\ell)}.$ Notice that for fixed $\ell$, these bounds are polynomial
in $k$.
One can also check whether $S$ is non-empty,
as well as compute a finite set of  sample points meeting every 
connected component of $S$ in time $k^{O(\ell)}$ \cite{Barvinok,GP}.
However, no algorithm with similar complexity is known for computing any of 
the individual Betti numbers of $S$ (for instance, the number
of connected components). The best known algorithm for computing all
the Betti numbers of $S$ has complexity $k^{2^{O(\ell)}}$ \cite{Basu05}.

Here, and elsewhere in  this  paper the Betti number, $b_i(S)$, is  the 
dimension of the simplicial homology group, $H_i(S,\Q),$ 
in case $S$ is closed and bounded. 
If $S$ is a closed but not necessarily bounded semi-algebraic set,
$b_i(S)$ is the dimension of $H_i(S \cap \overline{B_k(0,r)},\Q)$, 
for sufficiently large $r > 0$ (here and in the rest of the paper, 
$B_k(0,r)$ denotes the open ball of radius $r$ in $R^k$ centered 
at the origin, and 
$\overline{X}$ denotes the closure of a semi-algebraic set $X$). 
It is easy to see that $b_i(S)$ is well-defined and
we denote by 
\[
\chi(S) = \sum_{i=0}^{k} (-1)^i b_i(S)
\]
the Euler-Poincar\'e characteristic of $S$.

In this paper we describe an algorithm for computing the Euler-Poincar\'e
characteristic of $S$, whose complexity is $k^{O(\ell)}$. Our algorithm
relies on an efficient algorithm for computing the Euler-Poincar\'e
characteristic of the realizations of all realizable sign conditions
of a family of polynomials described in \cite{BPR04} and uses different
techniques than those used in \cite{Barvinok, GP}.

The main result of this paper is the following.\\
\noindent {\bf Main Result:} 
We present an algorithm (\ref{alg:basic} in 
\ref{sec:general})
which given a set of $\ell$ polynomials,
${\cal P} = \{P_1,\ldots,P_\ell\} \subset \R[X_1,\ldots,X_k],$
with ${\rm deg}(P_i) \leq 2, 1 \leq i \leq \ell,$
computes 
the Euler-Poincar\'e characteristic, $\chi(S)$,
where $S$ is the set defined by $P_1 \leq 0,\ldots,P_\ell \leq 0$.
The complexity of the algorithm  is
$ k^{O(\ell)}.$
If the coefficients of the polynomials in
${\cal P}$ are integers  of bitsizes bounded by
$\tau$, then the bitsizes of the integers
appearing in the intermediate computations and the output
are bounded by $\tau k^{{O(\ell)}}.$

The rest of the paper is organized as follows. In  \ref{sec:prelim},
we describe some mathematical and algorithmic results we will need for
our algorithm. We also include a brief introduction to spectral sequences
since they play a motivating role in the design of the main algorithm
described in this paper. In  \ref{sec:homogeneous}, we describe an
algorithm for computing the Euler-Poincar\'e characteristic of a set
defined by homogeneous quadratic inequalities.  Finally, in 
\ref{sec:general} we describe our algorithm for  the general (inhomogeneous)
case. 

For referring to well known results in real algebraic geometry we sometime
use reference \cite{BPR03} as a useful source.
\section{Preliminaries}
\label{sec:prelim}
In this section we describe some mathematical and algorithmic results we will
require in the rest of the paper.

\subsection{Definition of the Euler-Poincar\'e Characteristic}
For the purposes of our algorithm, it will be useful to define 
Euler-Poincar\'e characteristic for locally closed semi-algebraic sets.
We do this in terms of the Borel-Moore homology groups of such sets (defined
below).
This definition agrees with the definition of Euler-Poincar\'e
characteristic stated earlier for closed and bounded semi-algebraic sets.
They may be distinct for semi-algebraic sets which are closed but not 
bounded.

The simplicial homology groups  of a pair of
closed and bounded semi-algebraic sets $T \subset S \subset \R^k$ are
defined as follows.
Such a pair of closed, bounded semi-algebraic sets can be
triangulated \cite{BPR03} using  a pair of simplicial complexes $(K,A)$,
where $A$ is a sub-complex of $K$.
The  $p$-th simplicial homology group
of the pair
$(S,T)$, $H_p(S,T)$, is $H_p(K,A)$.
The dimension of $H_p(S,T)$ as a $\Q$-vector space is called the
  $p$-th Betti number of the pair
$(S,T)$ and denoted $b_p(S,T)$.
The  Euler-Poincar\'e characteristic
of the pair $(S,T)$ is
$$\chi(S,T) = \sum_i (-1)^i b_i(S,T) .$$

The $p$-th Borel-Moore homology group of $S\subset \R^k$,
denoted $H_p^{BM}(S)$, 
is defined in terms of the homology groups of a pair of 
closed and bounded semi-algebraic sets as follows.
For $r > 0$, let $S_r = S \cap B_k(0,r).$ 
Note that, for a locally closed semi-algebraic
set $S$, both $\overline{S_r}$ and $\overline{S_r}\setminus S_r$ are closed
and bounded and hence $H_p(\overline{S_r}, \overline{S_r}\setminus S_r)$ is
well defined.
Moreover, it is a consequence of Hardt's triviality theorem \cite{Hardt}
that the homology group $H_p(\overline{S_r}, \overline{S_r}\setminus S_r)$
is invariant for all sufficiently  large $r > 0$.
We define, 
$H_p^{BM}(S) = H_p(\overline{S_r}, \overline{S_r}\setminus S_r),$ for
$r >0$ sufficiently large, and it follows from the remark above that it
is well defined. 
The Borel-Moore homology groups are  invariant under semi-algebraic
homeomorphisms (see \cite{BCR}.
It also follows clearly from the definition that for a  closed and bounded
semi-algebraic set, the Borel-Moore homology groups coincide with the
simplicial homology groups.

For a locally closed semi-algebraic set $S$, we define the Borel-Moore
Euler-Poincar\'e characteristic by,
\[
\chi^{BM}(S) = \sum_{i=0}^k b_i^{BM}(S),
\]
where $b_i^{BM}(S)$ denotes the dimension of $H_i^{BM}(S,\Q).$
Note that if $S$ is closed and bounded, then $\chi^{BM}(S) = \chi(S).$

The Borel-Moore Euler-Poincar\'e characteristic has the following additive 
property.

\begin{proposition}
\index{Additivity of Euler-Poincar\'e characteristic}
\label{6:prop:additivity}
Let $X,X_1$ and $X_2$ be locally closed semi-algebraic
sets such that  $$X_1\cup X_2=X, X_1\cap X_2=\emptyset.$$  Then
$$\chi^{BM}(X)=\chi^{BM}(X_1)+\chi^{BM}(X_2).$$
\end{proposition}
\begin{proof}
This is classical 
(see for example, Proposition 2.6 in \cite{BPR04} for a proof).
\end{proof}

\subsection{Sign Conditions and their realizations}
A {\em sign condition} is an element of $\{0,1,- 1\}$.
We define
\[\s(x) =
\begin{cases}
0& {\rm if\ and\ only\ if\ }\ x=0\cr
1& {\rm if\ and\ only\ if\ }\ x> 0\cr
-1& {\rm if\ and\ only\ if\ }\ x < 0
\end{cases}
\]

Let $Z \subset \R^k$ be a locally closed semi-algebraic set and
let ${\mathcal P}= \{ P_1,\ldots,P_s \}$ 
be a finite subset of $\R[X_1,\ldots,X_k]$.
A  {\em sign condition}   on
${\mathcal P}$ is an element of $\{0,1,- 1\}^{\mathcal P}.$

The  realization of the  sign condition
${\sigma}$ on  ${Z}$  is
$$\displaylines{
\RR(\sigma,Z)= \{x\in Z\;\mid\;
 \bigwedge_{P \in {\cal P}} \s({P}(x))=\sigma(P) \},
}
$$
and its Euler-Poincar\'e characteristic is denoted $\chi^{BM}(\sigma,Z).$

We denote  by ${\rm Sign}({\cal P},Z)$   the list of
$\sigma \in  \{0,1,- 1\}^{\cal P}$
such that $\RR(\sigma,Z)$ is non-empty.
We denote by $\chi^{BM}({\cal P},Z)$  the  list of
Euler-Poincar\'e characteristics
$\chi^{BM}(\sigma,Z)=\chi^{BM}(\RR(\sigma,Z))$ for
$\sigma \in {\rm Sign}({\cal P},Z)$.

We will use the following algorithm for computing the list
$\chi^{BM}({\cal P},Z)$ described in \cite{BPR04}. We describe here the
input, output and complexity of the algorithm.

\begin{algorithm}{13:alg:eulerdet}
[Euler-Poincar\'e Characteristic of Sign Conditions]
\item 
an algebraic set $Z=\ZZ(Q,\R^k) \subset\R^k$  and a
finite list ${\cal P}=P_1,\ldots,P_s$ of polynomials  in
$\R[X_1,\ldots,X_k]$.
\item
the list $\chi^{BM}({\cal P},Z)$.
\end{algorithm}

\noindent\textsc{Complexity:}
Let $k'$ be the dimension of $Z$, $d$ a bound on the degree of $Q$ and the
elements
of ${\cal P}$
and $s=\#({\cal P}))$.
The number of arithmetic operations is
$$s^{k'+1}O(d)^k+ s^{k'}((k'\log_2(s)+k\log_2(d))d)^{O(k)}.$$
The algorithm also involves the inversion  matrices of
size $s^{k'}O(d)^k$ with integer coefficients.

If $\D=\Z,$ and the bitsizes
of the coefficients of the polynomials are bounded by
 $\tau$, then the bitsizes of the integers
appearing in the intermediate computations and the output
are bounded by
$\tau((k'\log_2(s)+k\log_2(d))d)^{O(k)}$.

\subsection{Infinitesimals}
In our algorithms we will use infinitesimals in order to
ensure that the set we are dealing with is bounded.
To ensure this we will extend the ground field $\R$ 
to $\R\langle \varepsilon\rangle$,  the real closed field of algebraic
Puiseux series in $\varepsilon$ with coefficients in $\R$.
The sign of a Puiseux series in $\R\langle \varepsilon\rangle$
agrees with the sign of the coefficient
of the lowest degree term in
$\varepsilon$. 
This induces a unique order on $\R\langle \varepsilon\rangle$ which
makes $\varepsilon$
infinitesimal: $\varepsilon$ is positive and smaller than
any positive element of $\R$.

If $\R'$ is a real closed field containing $\R$, then
given a semi-algebraic set
$S$ in ${\R}^k$, we denote the {\em extension}
of $S$ to $\R'$ by $\E(S,\R')$.
$\E(S,\R')$  is
the semi-algebraic subset of ${ \R'}^k$ defined by the same
quantifier free formula that defines $S$.
The set $\E(S,\R')$ is well defined (i.e. it only depends on the set
$S$ and not on the quantifier free formula chosen to describe it).
This is an easy consequence of the Tarski-Seidenberg 
transfer principle (see for example Section 2.4.1 in \cite{BPR03}).

\subsection{Spectral Sequences}
\label{subsec:sepctral}
For the benefit of the readers 
we include  a brief introduction to the theory of spectral sequences
pointing to \cite{Bredon, Mcleary} for more details.
  
A {\em spectral sequence} is a sequence of bigraded complexes
$(E_r, d_r: E^{p,q}_r \rightarrow E^{p+r,q-r+1}_r)$ such that 
the complex $E_{r+1}$ is obtained from $E_r$ by taking its homology
with respect to $d_r$ (that is $E_{r+1} = H_{d_r}(E_r)$).

{\small
\begin{figure}[hbt] 
\begin{center}
\begin{picture}(0,0)%
\includegraphics{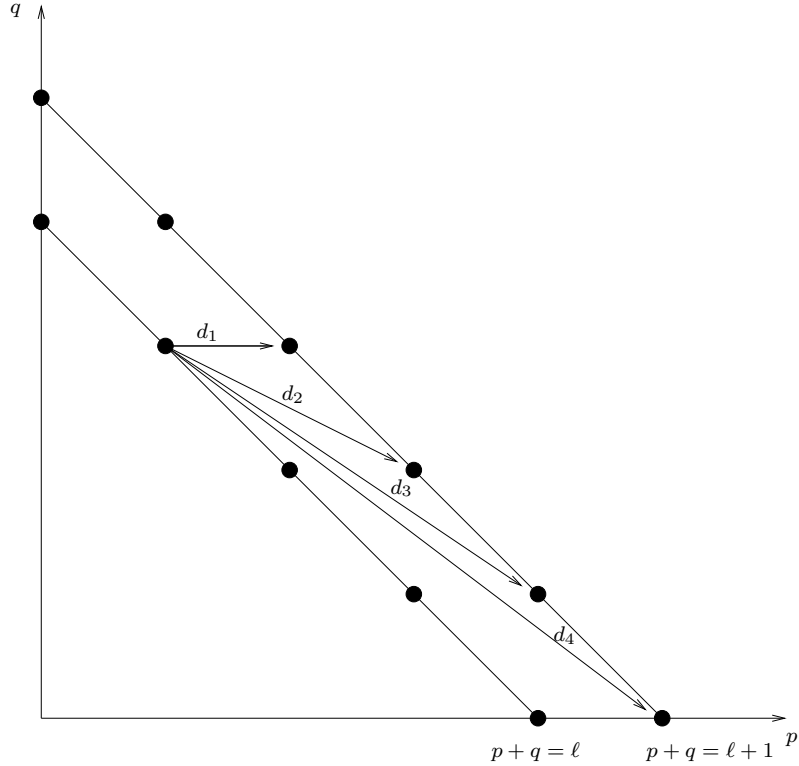}%
\end{picture}%
\setlength{\unitlength}{2565sp}%
\begingroup\makeatletter\ifx\SetFigFont\undefined%
\gdef\SetFigFont#1#2#3#4#5{%
  \reset@font\fontsize{#1}{#2pt}%
  \fontfamily{#3}\fontseries{#4}\fontshape{#5}%
  \selectfont}%
\fi\endgroup%
\begin{picture}(7512,7414)(601,-7394)
\put(6751,-7336){\makebox(0,0)[lb]{\smash{\SetFigFont{8}{9.6}{\familydefault}{\mddefault}{\updefault}{$p + q = \ell+1$}%
}}}
\put(5251,-7336){\makebox(0,0)[lb]{\smash{\SetFigFont{8}{9.6}{\familydefault}{\mddefault}{\updefault}{$p + q = \ell$}%
}}}
\put(8101,-7186){\makebox(0,0)[lb]{\smash{\SetFigFont{8}{9.6}{\familydefault}{\mddefault}{\updefault}{$p$}%
}}}
\put(601,-136){\makebox(0,0)[lb]{\smash{\SetFigFont{8}{9.6}{\familydefault}{\mddefault}{\updefault}{$q$}%
}}}
\put(2401,-3286){\makebox(0,0)[lb]{\smash{\SetFigFont{8}{9.6}{\familydefault}{\mddefault}{\updefault}{$d_1$}%
}}}
\put(3226,-3886){\makebox(0,0)[lb]{\smash{\SetFigFont{8}{9.6}{\familydefault}{\mddefault}{\updefault}{$d_2$}%
}}}
\put(4276,-4786){\makebox(0,0)[lb]{\smash{\SetFigFont{8}{9.6}{\familydefault}{\mddefault}{\updefault}{$d_3$}%
}}}
\put(5851,-6211){\makebox(0,0)[lb]{\smash{\SetFigFont{8}{9.6}{\familydefault}{\mddefault}{\updefault}{$d_4$}%
}}}
\end{picture}
\caption{$d_r: E_r^{p,q} \rightarrow E_r^{p+r, q- r +1}$}
\label{fig:spectral}
\end{center}
\end{figure}
}

There are two spectral sequences,
${'E}_*^{p,q},{''E}_*^{p,q}$,  (corresponding to taking row-wise or
column-wise filtrations respectively) 
associated with a
double complex ${C}^{\bullet,\bullet},$ 
which  will be important for us. 
Both of these converge to $H^*({\rm Tot}^{\bullet}(C^{\bullet,\bullet})).$
This means that the homomorphisms $d_r$ are eventually zero, and hence the
spectral sequences stabilize, and
\[
\bigoplus_{p+q = i} {'E}_{\infty}^{p,q} \cong 
\bigoplus_{p+q = i} {''E}_{\infty}^{p,q} \cong 
H^i({\rm Tot}^{\bullet}(C^{\bullet,\bullet})),
\]
for each $i \geq 0$.

The first terms of these are
${'E}_1 = H_{\delta}(C^{\bullet,\bullet}), 
{'E}_2 = H_dH_{\delta}(C^{\bullet,\bullet})$, and
${''E}_1 = H_d (C^{\bullet,\bullet}), 
{''E}_2 = H_\delta H_d (C^{\bullet,\bullet})$.

In particular, assuming that the complex $C^{\bullet,\bullet}$ is bounded
in both directions, we have that,
\begin{proposition}
\label{prop:euler}
$$
\begin{array}{lll}
\sum_{i\geq 0} (-1)^i \dim (H^i({\rm Tot}^{\bullet}(C^{\bullet,\bullet})))
&=& \sum_{p,q \geq 0} (-1)^{p+q} \dim ({'E}_2^{p,q})\\
&=& \sum_{p,q \geq 0} (-1)^{p+q} \dim ({''E}_2^{p,q}).
\end{array}
$$
\end{proposition}

\subsection{Leray Spectral Sequence of a map}
The Leray spectral sequence of a proper map,
\[
f: A \longrightarrow B,
\]
is a classical tool in
algebraic topology which relates the cohomology of the space $A$, with
those of the space $B$ and of the fibers of the map $f$. 
Its most common use is in the theory of sheaf cohomology \cite{Bredon}. 
We will need it in a very
special situation where the sets $A$ and $B$ are compact semi-algebraic
sets and $f$ a continuous semi-algebraic map. In this special situation it
is possible to define the Leray spectral sequence in terms of triangulations,
which we do below.

Consider a semi-algebraic continuous map,
$f: A \longrightarrow B$, where $A$ and $B$ are compact semi-algebraic sets.
Moreover, let $h: \Delta \longrightarrow B$ be a semi-algebraic triangulation
of $B$, and let ${\mathcal H}(A)$ denote a cell-complex, such that $A$ is the union
of the cells in ${\mathcal H}(A)$,
and such that for any simplex $\sigma \in \Delta$,
$A_\sigma = f^{-1}(\overline{h(\sigma)})$ is a subcomplex of ${\mathcal H}(A)$ (
where $\overline{X}$ denotes the topological closure of $X$).

Then, the Leray spectral sequence of $f$ is isomorphic to the spectral
sequence (corresponding to the column-wise filtration) associated to the
double complex ${C}^{\bullet,\bullet}$ defined as follows:
\[ 
C^{p,q} = 
\bigoplus_{\sigma \;\mbox{a $p$-simplex in $\Delta$}} C^q(A_\sigma),
\]
where $C^q(A_\sigma)$ denotes the vector space of $q$-co-chains of
the complex $A_\sigma$. The horizontal and the vertical differentials
are the obvious ones (see \cite{Bredon}).
The spectral sequence associated to the double complex defined above converges
to the co-homology of $A$.

\section{The basic homogeneous case}
\label{sec:homogeneous}
Let ${\cal P} = \{P_1,\ldots,P_\ell\} \subset \R[X_0,X_1,\ldots,X_k]$
be a set of homogeneous quadratic polynomials, and
let $S$ be the basic closed semi-algebraic set defined on the unit sphere
$S^k \subset \R^{k+1}$ by the inequalities,
\[
P_1 \leq 0,\ldots, P_\ell \leq 0.
\]
We denote by $S_i$ the subset of $S^k$ defined by $P_i \leq 0$. 
Then, $S = \cap_{i=1}^{\ell} S_i$. 
For $J \subset \{1,\ldots,\ell\}$, we denote by
$S^J = \cup_{j \in J}S_j.$
The following equality
is a consequence of the exactness of the Mayer-Vietoris sequence.
\begin{lemma}
\label{lem:MV}
\[
\chi(S) = \sum_{J \subset \{1,\ldots,\ell\}} (-1)^{\#(J)+1} \chi(S^J).
\]
\end{lemma}
\begin{proof}
In case $\ell=2$, this is a direct consequence of the exactness of
Mayer-Vietoris sequence (see for example, \cite{BPR03}, Corollary 6.28).
The general case follows from an easy induction. 
\end{proof}

Thus, in order to compute $\chi(S)$ it suffices to compute
$\chi(S^J)$ for each $J \subset \{1,\ldots,\ell\}$.

\subsection{Topology of sets defined by quadratic constraints}
\label{sec:top_quadratic}
We first recall some facts about topology of sets defined by
quadratic inequalities \cite{Agrachev}.
Let 
$P_1,\ldots,P_{s}$ be homogeneous quadratic polynomials in
$\R[X_0,\ldots,X_k]$. 

We denote by $P = (P_1,\ldots,P_s): \R^{k+1} \rightarrow \R^s$,
the map defined by the polynomials
$P_1,\ldots,P_s.$  
Let
$$
A = \bigcup_{1 \leq i \leq s}\{ x \in S^k \mid   P_i(x) \leq   0 \}.
$$ 

Let 
$$
\Omega = \{\omega \in R^{s} \mid  |\omega| = 1, \omega_i \leq 0, 1 \leq i \leq s\}.
$$

For $\omega \in \Omega$ we denote by ${\omega}P$ the quadratic form
defined by 
\[
{\omega}P = \sum_{i=1}^{s} \omega_i P_i.
\]

Let $B \subset \Omega \times S^k$ be the set defined by,
\[
B = \{ (\omega,x)\mid \omega \in \Omega, x \in S^k \;\mbox{and} \; 
{\omega}P(x) \geq 0\}.
\]

We denote by $\phi_1: B \rightarrow \Omega $ and 
$\phi_2: B \rightarrow S^k$ the two projection maps. 

\begin{diagram}
&&B&& \\
& \ldTo^{\phi_1} &&\rdTo^{\phi_2} & \\
\Omega &&&& S^k
\end{diagram}

The following was proved by Agrachev \cite{Agrachev}.
With the notation developed above,
\begin{proposition}
\label{prop:homotopy2}
The map $\phi_2$ gives a homotopy equivalence between $B$ and 
$\phi_2(B) = A$.
\end{proposition}

\begin{proof}
We first prove that $\phi_2(B) = A.$
If $x \in A,$ 
then there exists some $i, 1 \leq i \leq s,$ such that
$P_i(x) \leq 0.$ Then for $\omega = (-\delta_{1i},\ldots,-\delta_{si})$
(where $\delta_{ij} = 1$ if $i=j$, and $0$ otherwise),
we see that $(\omega,x) \in B.$
Conversely,
if $x \in \phi_2(B),$ then there exists 
$\omega = (\omega_1,\ldots,\omega_s) \in \Omega$ such that,
$\sum_{i=1}^s \omega_i P_i(x) \geq 0$. Since, 
$\omega_i \leq 0, 1\leq i \leq s,$ and not all $\omega_i = 0$,
this implies that $P_i(x) \leq 0$ for
some $i, 1 \leq i \leq s$. This shows that $x \in A$.

For $x \in \phi_2(B)$, the fibre 
$$
\phi_2^{-1}(x) = \{ (\omega,x) \mid  
 \omega \in \Omega \;\mbox{such that} \;  {\omega}P(x) \geq 0\},
$$
can be identified with a 
non-empty subset of $\Omega$ defined by a single linear inequality. 
From convexity considerations, all 
such fibres can clearly be retracted to their center of mass continuously,
proving the first half of the proposition.
\end{proof}

For any  quadratic form $Q$, we will denote by ${\rm index}(Q)$, the number of
negative eigenvalues of the symmetric matrix of the corresponding bilinear
form, that is of the matrix $M$ such that,
$Q(x) = \langle Mx, x \rangle$ for all $x \in R^{k+1}$. We will also
denote by $\lambda_i(Q), 0 \leq i \leq k$, the eigenvalues of $Q$, in non-decreasing order, i.e.
\[ \lambda_0(Q) \leq \lambda_1(Q) \leq \cdots \leq \lambda_k(Q).
\]

Given a quadratic map $P = (P_1,\ldots,P_s): \R^{k+1} \rightarrow \R^s,$
and $0 \leq j \leq k$, 
we denote by 
\[
\Omega_j = \{\omega \in \Omega \;  \mid \;  \lambda_j({\omega}P) \geq 0 \}.
\]
For notational convenience, $\Omega_{-1}$ will denote the empty set and
$\Omega_{k+1}$ the whole space $\Omega$.

It is clear that the $\Omega_j$'s induce a filtration of the space
$\Omega$, i.e.,
$
\Omega_0 \subset \Omega_1 \subset \cdots \subset \Omega_{k+1}.
$

Agrachev \cite{Agrachev} showed that the Leray spectral sequence of the map
$\phi_1$ (converging to the cohomology $H^*(B) \cong H^*(A)$),
has as its $E_2$ terms,
\begin{equation}
\label{eqn:agrachev}
E_2^{pq} = H^p(\Omega_{k-q},\Omega_{k-q-1}).
\end{equation}

\label{obs:sphere}
This follows from the fact that the fibre of the map $\phi_1$ over a point 
$\omega \in \Omega_{j}\setminus \Omega_{j-1}$ has the homotopy type
of a sphere of dimension $k-j$. 
To see this notice that for
$\omega \in  \Omega_{j}\setminus \Omega_{j-1}$,
$\lambda_0({\omega}P),\ldots, \lambda_{j-1}({\omega}P) < 0.$ Moreover,
letting $Y_0({\omega}P),\ldots,Y_k({\omega}P)$ be 
an orthonormal basis consisting of the
eigenvectors of ${\omega}P$, we have that 
$\phi_1^{-1}(\omega)$ is the subset of $S^k$ defined by,
$$
\displaylines{
\sum_{i=0}^{k} \lambda_i({\omega}P)Y_i({\omega}P)^2 \geq  0, \cr
\sum_{i=0}^{k} Y_i({\omega}P)^2 = 1.
}
$$

Since, $\lambda_i({\omega}P) < 0, 0 \leq i < j,$ it follows that
for $\omega \in \Omega_{j}\setminus \Omega_{j-1},$
$\phi_1^{-1}(\omega)$  is homotopy equivalent to the
$(k-j)$-dimensional sphere defined by setting
$Y_0({\omega}P) = \cdots = Y_{j-1}({\omega}P) = 0$ on the sphere defined by
$\sum_{i=0}^{k}Y_i({\omega}P)^2 = 1.$

The following proposition relates
the Euler-Poincar\'e characteristic of the set
$A$ with those of $\Omega_j\setminus \Omega_{j-1}, \; 0 \leq j \leq k+1.$
\begin{proposition}
\label{prop:chi1}
\[
\chi(A) = 
\chi^{BM}(A) = \sum_{j=0}^{k+1} \chi^{BM}(\Omega_{j}\setminus \Omega_{j-1})
(1 + (-1)^{(k-j)}).
\]
\end{proposition}

\begin{proof}
Notice that the sets $\Omega_j \setminus \Omega_{j-1}$ are locally closed,
and the fibre over a point $\omega \in \Omega_j \setminus \Omega_{j-1}$
is compact and has the homotopy type of a $(k-j)$-dimensional sphere.
Now consider a sufficiently fine triangulation of $\Omega$, which respects
the filtration $\Omega_0 \subset \cdots \subset \Omega_{k+1}$, and such that
over each simplex $\sigma$ of the triangulation lying in 
$\Omega_j \setminus \Omega_{j-1}$,
$\phi_1^{-1}(\sigma)$ is homotopy equivalent to $\sigma \times S^{k-j}$.
The Euler-Poincar\'e characteristic of a $(k-j)$-dimensional sphere,
$S^{k-j}$, is
equal to $1 + (-1)^{(k-j)}.$
The proposition now follows from the additivity property of the Borel-Moore
Euler-Poincar\'e characteristic and  \ref{prop:homotopy2}.

\end{proof}

Since \ref{prop:chi1} is central to the algorithm presented in this
paper, we include a different proof below which uses the spectral sequence
\ref{eqn:agrachev}.
First note that by \ref{6:prop:additivity},
\[
\chi^{BM}(\Omega_{j}\setminus \Omega_{j-1}) = \chi(\Omega_j) - \chi(\Omega_{j-1}).
\]
It follows from the convergence of the spectral sequence in 
\ref{eqn:agrachev} and \ref{prop:euler} that,
$$\displaylines{
\begin{array}{lll}
\chi(A) &=& \sum_{p+q=i} (-1)^i \dim(E_2^{p,q}) \\
 &=& \sum_{p+q = i} (-1)^i \dim H^p(\Omega_{k-q},\Omega_{k-q-1})\\
&=& \sum_{0 \leq q \leq k+1} 
\sum_{0 \leq p \leq k} (-1)^{p+q} \dim (H^p(\Omega_{k-q},\Omega_{k-q-1})) \\
&=& \sum_{0 \leq q \leq k+1} (-1)^q 
\sum_{0 \leq p \leq k} (-1)^{p} \dim (H^p(\Omega_{k-q},\Omega_{k-q-1}))
\end{array}
}
$$
Now, from the exact sequence of the pair $(\Omega_{k-q},\Omega_{k-q-1})$,
namely,
\[
\cdots \rightarrow H^{i-1}(\Omega_{k-q}) \rightarrow H^{i-1}(\Omega_{k-q-1}) 
\rightarrow H^i(\Omega_{k-q},\Omega_{k-q-1})\rightarrow
H^{i}(\Omega_{k-q}) \rightarrow \cdots,
\]
we get  that,
\[
\sum_{i \geq 0} (-1)^i( \dim(H^i(\Omega_{k-q-1})) - \dim (H^i(\Omega_{k-q}))
+  \dim(H^i(\Omega_{k-q},\Omega_{k-q-1})) = 0,
\]
which yields
\[\sum_{0 \leq p \leq k} (-1)^{p} \dim (H^p(\Omega_{k-q},\Omega_{k-q-1}))
= \chi(\Omega_{k-q}) - \chi(\Omega_{k-q-1}).
\]

Thus, the previous sum
$$\displaylines{
\begin{array}{lll}
&=&\sum_{0 \leq q \leq k+1} (-1)^q (\chi(\Omega_{k-q}) - \chi(\Omega_{k-q-1})) \\
&=& \sum_{0 \leq j \leq k+1}(-1)^{k+1 - j}
(\chi(\Omega_{j - 1}) - \chi(\Omega_{j-2})) \\
&=& \sum_{0 \leq j \leq k+1}(-1)^{k+1-j}\chi(\Omega_{j-1})
 -  \sum_{0 \leq j \leq k+1}(-1)^{k+1-j}\chi(\Omega_{j-2})\\
&=& \sum_{0 \leq j \leq k}(-1)^{k-j}\chi(\Omega_{j})
 -  \sum_{0 \leq j \leq k-1}(-1)^{k+1-j}\chi(\Omega_{j})\\
&=& \sum_{0 \leq j \leq k}(-1)^{k-j}\chi(\Omega_{j})
 +  \sum_{0 \leq j \leq k-1}(-1)^{k-j}\chi(\Omega_{j})\\
&=& \chi(\Omega_k) - 2\chi(\Omega_{k-1}) + 2\chi(\Omega_{k-2}) + \cdots + 
(-1)^k 2\chi(\Omega_0)\\
&=&  \sum_{0 \leq j \leq k+1} (\chi(\Omega_j) - \chi(\Omega_{j-1}))(1 + (-1)^{k-j}) \\
&=&  \sum_{0 \leq j \leq k+1} \chi^{BM}(\Omega_j\setminus\Omega_{j-1})(1 + (-1)^{k-j}).
\end{array}
}
$$
\Qed

Let $Z= (Z_1,\ldots,Z_s)$ be variables and let $M(Z)$ be the symmetric
matrix corresponding to the quadratic form 
$Z\cdot P = Z_1P_1 +\cdots+Z_sP_s$. The entries
of $M(Z)$ depend linearly on $Z$. Let
\[
F(Z,T) = \det(M(Z) + T\cdot I_{k+1})= T^{k+1} + C_{k}T^{k} + \cdots+ 
C_0,
\]
where each $C_i \in \R[Z_1,\ldots,Z_s]$ is a polynomial of degree
at most $k+1$.

It follows from the well known Descarte's rule of signs 
(see for example, Remark 2.42 in \cite{BPR03}) that for any 
$z \in \Omega$,
${\rm index}(zP)$ is equal to the number of sign variations in the
sequence $C_0(z),\ldots,C_{k}(z),+1$. Thus, the signs of the polynomials
$C_0,\ldots,C_k$ determine the index of $zP$. For 
$\sigma \in \{0,+1,-1\}^{{\cal C}}$
a sign condition on the family ${\cal C} = \{C_0,\ldots,\C_k\}$, let
$n(\sigma)$ denote the number of sign variations in the sequence,
$\sigma(C_0),\ldots,\sigma(C_k),+1.$
Let $\SIGN({\cal C},\Omega)$ be the set of sign conditions realized
by the family ${\cal C}$ on $\Omega$.
The following proposition is an immediate consequence of 
\ref{prop:chi1} and the additivity of the Euler-Poincar\'e
characteristic.

\begin{proposition}
\label{prop:chi2}
\[
\chi(A) = \chi^{BM}(A) = \sum_{\sigma \in \SIGN({\cal C},\Omega)} \chi^{BM}(\RR(\sigma,\Omega))
\cdot (1 + (-1)^{(k - n(\sigma))}).
\]
\end{proposition}

Before proceeding further we discuss a small example.

\begin{example}
Let $\ell = 2, k = 2$ and $P = (P_1,P_2): \R^3 \rightarrow \R^2$ be the
quadratic map with,
$$
\displaylines{
P_1 = X_0^2 + X_1^2 - X_2^2, \cr
P_2 = X_0^2 - X_1^2 - X_2^2.
}
$$

In this example,
$$
\Omega = 
\{(\omega_1,\omega_2)\; \mid \; \omega_1^2 + \omega_2^2 = 1, 
\omega_1,\omega_2 \leq 0 \}
$$
consists of the arc of the unit circle in the third quadrant.

Also,
$$
\begin{array}{lll}
A  &=& \{ x \in S^2 \mid   P_1(x) \leq 0 \vee P_2(x) \leq 0 \} \cr
  &=& \{ (x_0,x_1,x_2) \in \R^3 \mid  x_0^2 + x_1^2 + x_2^2 = 1, 
x_1^2 + x_2^2 \geq 1/2\}.
\end{array}
$$
The set $A$ in this example clearly has the homotopy type of a circle,
and hence,
\begin{equation}
\label{eqn:example}
\chi(A) = 0.
\end{equation}

Now, for $\omega = (\omega_1,\omega_2) \in \Omega,$ 
$$
\begin{array}{lll}
\omega P &=& \omega_1 P_1 + \omega_2 P_2 \cr
         &=& (\omega_1 + \omega_2)X_0^2 + (\omega_1 - \omega_2)X_1^2 
             - (\omega_1 + \omega_2)X_2^2.
\end{array}
$$

Following notations introduced above,
$$
\displaylines{
F(Z_1,Z_2,T) = (Z_1 + Z_2 + T)(Z_1 - Z_2 + T)(-Z_1 - Z_2 + T) \cr
             = T^3 + (Z_1 - Z_2)T^2 - (Z_1 + Z_2)^2 T + (Z_1+Z_2)^2(Z_2 - Z_1).
}
$$

Thus, in this example,
$$
\displaylines{
C_0(Z_1,Z_2) = (Z_1+Z_2)^2(Z_2 - Z_1),\cr
C_1(Z_1,Z_2) = - (Z_1 + Z_2)^2, \cr
C_2(Z_1,Z_2) = Z_1 - Z_2.
}
$$

There are three realizable sign conditions on
${\cal C} = \{C_0,C_1,C_2,+1\}$ on $\Omega$. They are,
$$
\displaylines{
\sigma_1 = (-,-,+,+), \cr
\sigma_2 = (0,-,0,+), \cr
\sigma_3 = (+,-,-,+).
}
$$

We have,
$$
\displaylines{
n(\sigma_1) = 1, \cr
n(\sigma_2) = 1, \cr
n(\sigma_3) = 2.
}
$$

The realizations $\RR(\sigma_1,\Omega)$ and $\RR(\sigma_3,\Omega)$
are each homeomorphic to $[0,1)$ while $\RR(\sigma_2,\Omega)$ is a point.
Thus, 
\[
\chi^{BM}(\sigma_1,\Omega) = \chi^{BM}(\sigma_3,\Omega) = 0,
\]
while
\[
\chi^{BM}(\sigma_2,\Omega) = 1.
\]

Finally, we have
$$
\begin{array}{ccl}
\sum_{j=1}^{3}\chi^{BM}(\sigma_j,\Omega)(1 - (-1)^{2-n(\sigma_j)})&=&
0(1 + (-1)^1) + 1(1 + (-1)^1) \\
&& +\; 0(1 + (-1)^2)\\
&=& 0,
\end{array}
$$
which agrees with \ref{eqn:example}.
\Qed
\end{example}

We are now in a position to describe an algorithm for computing the 
Euler-Poincar\'e characteristic of a union of sets, each  defined by
a homogeneous quadratic inequality.
In the algorithm we will use the following notation.
Given two finite families of polynomials, ${\cal P} \subset {\cal P}'$,
and $\sigma \in \{0,+1,-1\}^{{\cal P}}, \sigma' \in \{0,+1,-1\}^{{\cal P}'}$,
we say that $\sigma \prec \sigma'$ iff
for all $P \in {\cal P},$ $\sigma(P) = \sigma'(P)$.

\begin{algorithm} {alg:union}
[Euler-Poincar\'e characteristic of a union]
\item A set of quadratic forms
$\{P_1,\ldots,P_s\}\subset  \R[X_0,\ldots,X_k].$ 
\item 
$\chi(A)$, where $A$ is the set defined on the unit sphere $S^k \subset 
\R^{k+1}$ by the formula
$$
P_1 \leq 0 \vee \cdots \vee P_\ell \leq 0.
$$
\item
Let $P = (P_1,\ldots,P_s)$.
Let $Z= (Z_1,\ldots,Z_s)$ be variables and let $M(Z)$ be the symmetric
matrix corresponding to the quadratic form $Z\cdot P$. 
Compute the polynomials $C_i \in \R[Z_1,\ldots,Z_s]$ by computing
the following determinant.
\[
\det(M(Z) + T\cdot I_k)= T^{k+1} + C_{k}T^{k} + \cdots+ 
C_0.
\]
\item
Compute $\chi^{BM}({\cal C},\Omega)$ as follows.
Call \ref{13:alg:eulerdet} with input 
${\cal C}' = {\cal C} \cup \{Z_1,\ldots,Z_s\}$ and
$Q = Z_1^2 +\cdots+ Z_s^2 - 1.$
The output is the list 
\[\chi^{BM}({\cal C}',\ZZ(Q,\R^k)).
\]

For each $\sigma \in \{0,+1,-1\}^{{\cal C}}$, such that there exists
$\sigma' \in \SIGN({\cal C'}, \ZZ(Q,\R^k))$ with 
$\sigma \prec \sigma'$ and $\sigma'(Z_j) \in \{0,-1\}$ for 
$1 \leq j \leq s,$
compute
$$
\displaylines{
\chi^{BM}(\sigma,\Omega) = 
\sum_{\sigma',\sigma \prec \sigma',\sigma'(Z_j) \in \{0,-1\},1 \leq j \leq s}
\chi^{BM} (\sigma',\ZZ(Q,\R^k)).
}
$$

\item
Output 
\[
\chi(A) = 
\sum_{\sigma \in \SIGN({\cal C},\Omega)} \chi^{BM}(\RR(\sigma,\Omega))
\cdot (1 + (-1)^{(k - n(\sigma))}).
\]
\end{algorithm}

\noindent\textsc{Proof of Correctness:}
The correctness of the algorithm is a consequence of 
\ref{prop:chi2} and the correctness of \ref{13:alg:eulerdet}.
\Qed

\noindent\textsc{Complexity Analysis:}
The complexity of the algorithm is  $k^{O(s)}$ using the complexity of 
\ref{13:alg:eulerdet}.
\Qed

We are now in a position to describe the algorithm for computing the
Euler-Poincar\'e characteristic in the basic, homogeneous case.

\begin{algorithm}{alg:homogeneous} 
[The basic homogeneous case]
\item A set of quadratic forms
$\{P_1,\ldots,P_\ell\}\subset  \R[X_0,\ldots,X_k].$ 
\item 
$\chi(S)$, where $S$ is the set defined on the unit sphere $S^k \subset 
\R^{k+1}$ by the inequalities,
$$
P_1 \leq 0,\ldots, P_\ell \leq 0.
$$
\item
For each subset $J \subset \{1,\ldots,\ell\}$ do the following.

\item
Compute using \ref{alg:union} $\chi(S^J).$

\item
Output 
\[
\chi(S) = \sum_{J \subset \{1,\ldots,\ell\}} (-1)^{\#(J)+1} \chi(S^J).
\]
\end{algorithm}

\noindent\textsc{Proof of Correctness:}
The correctness of the algorithm is a consequence of Lemma \ref{lem:MV}
and the correctness of  \ref{alg:union}.
\Qed
\noindent\textsc{Complexity Analysis:}
There are $2^{\ell}$ calls to  \ref{alg:union}. Using the
complexity analysis of  \ref{alg:union}, the complexity of
the algorithm is bounded by $k^{O(\ell)}.$
\Qed

\section{The General Case}
\label{sec:general}
Let ${\cal P} = \{P_1,\ldots,P_\ell\} \subset \R[X_1,\ldots,X_k]$ with
${\rm deg}(P_i) \leq 2, 1 \leq i \leq \ell,$ and 
let $S \subset \R^k$ be the basic semi-algebraic set defined by
$P_1 \leq 0, \ldots, P_\ell \leq 0$. 
Let $0 < \varepsilon$ be
an infinitesimal, and let 
$$
P_{\ell+1} = \varepsilon\sum_{j=1}^{k} X_j^2 - 1.
$$ 
Let $S' \subset \R\langle {\varepsilon}\rangle^{k}$ be the set defined
by 
$P_1 \leq 0,\ldots,P_{\ell+1} \leq 0.$

Denoting by ${P}_i^h$ the homogenization of ${P}_i$,
and ${S}^h \subset S^k$ the
set defined by, 
${P}^h_1 \leq 0,\ldots, {P}^h_\ell \leq 0, P^h_{\ell+1} \leq 0,$
on the unit sphere in $\R\langle {\varepsilon}\rangle^{k+1}$
we have,
\begin{proposition}
\label{prop:perturb2}
For $0 \leq i \leq k,$
$\chi(S) = \chi(S') = {1 \over 2}\chi({S}^h).$
\end{proposition}

\begin{proof}
Using the well known conic structure at infinity of semi-algebraic sets
(see for example Proposition 5.50, \cite{BPR03}) we have  that
for all sufficiently large $r > 0$, $S \cap \overline{B_k(0,r)}$ is 
a semi-algebraic deformation retract of $S$. Since $\eps$ is an infinitesimal, 
it follows that $S' = \E(S,\R\langle\eps\rangle) \cap 
\overline{B_k(0,\frac{1}{\eps})}$
is a semi-algebraic deformation retract of $\E(S,\R\langle\eps\rangle).$
This implies that  $\chi(S) = \chi(S')$.
 
To prove the second equality, first observe that $S'$ is bounded, and 
${S}^h$ is the projection from the origin 
of the set $1 \times S' \subset {1} \times \R\langle\varepsilon\rangle^k$
onto the unit sphere in $\R^{k+1}$. Since, ${S}'$ is bounded, 
the projection
does not intersect the equator and consists of two disjoint copies in the 
upper  and lower
hemispheres, and each copy is homeomorphic to $S'.$
\end{proof}

\begin{algorithm} {alg:basic}
[The general case]
\item A family of polynomials
${\cal P} = \{P_1,\ldots,P_\ell\}\subset  \R[X_1\ldots,X_k],$ with
${\rm deg}(P_i) \leq 2.$
\item 
$\chi(S)$, where $S$ is the set defined by
$$S = \bigcap_{P \in {\cal P}}
               \{x \in \R^{k}\; \mid \; P(x) \leq 0 \}.
$$
\item
Replace the family ${\cal P}$ by the family,
${\cal P}^h = \{ {P}^h_1,\ldots,{P}^h_\ell, P^h_{\ell+1} \}.$
\item
Using \ref{alg:homogeneous} compute 
$\chi({S}^h).$
\item
Output $\chi(S) = {1\over 2}\chi({S}^h).$
\end{algorithm}

\noindent\textsc
{Proof of Correctness:}
The correctness of  \ref{alg:basic} is a consequence of
\ref{prop:perturb2} and the
correctness of \ref{alg:homogeneous}.
\Qed

\noindent\textsc
{Complexity Analysis:}
The complexity of the algorithm is clearly $k^{O(\ell)}$
from complexity analysis of  \ref{alg:homogeneous}.
\Qed

\begin{remark}
{\em
In this paper we have described an algorithm for computing the Euler-Poincar\'e
characteristic of a basic closed semi-algebraic set defined by a 
constant number of
quadratic inequalities $P \leq 0, P \in {\cal P}$. 
It is straightforward to extend the algorithm
to the case of semi-algebraic sets defined by Boolean formulas without 
negations, whose atoms are of the form $P \geq 0$ or $P \leq 0$ for 
$P \in {\cal P}$,
using the technique used in \cite{B99} for reducing the
problem of computing Euler-Poincar\'e 
characteristic of such sets, to the basic closed case.
This reduction works perfectly well even in the quadratic situation
and does not worsen the complexity.
}
\end{remark}

\begin{acknowledge}
The author was supported in part by
an NSF Career Award 0133597 and a Sloan Foundation Fellowship.
\end{acknowledge}

\end{document}